\documentclass[preprint]{elsarticle}

\usepackage{amssymb,amsmath}
\usepackage{tikz}
\usetikzlibrary{patterns}

\journal{Journal of Computational Physics}
\begin{document}

\begin{frontmatter}



\title{Recursive Green's function method for multi-terminal nanostructures}


\author[ru,hi]{G.\ Thorgilsson}
\author[ru]{G.\ Viktorsson}
\author[ru]{S.I.\ Erlingsson}
\address[ru]{School of Science and Engineering, Reykjavik University, Menntavegi 1, IS-101 Reykjavik, Iceland}
\address[hi]{Science Institute, University of Iceland, Dunhaga 3, IS-107 Reykjavik, Iceland}
\begin{abstract}
We present and review an efficient method to calculate the retarded Green's function in multi-terminal nanostructures; which is needed in order
to calculate the conductance through the system and the local particle densities within it. The method uses the recursive Green's function method
after the discretized Hamilton matrix has been properly partitioned. We show that this method, the circular slicing scheme, 
can be modified to accommodate multi-terminal systems as well as the traditional two-terminal systems. Furthermore,
we show that the performance and robustness of the circular slicing scheme is \emph{on par} with other advanced methods and is well suited for large variety
of multi-terminal geometries. We end by giving an example of how the method can be used to calculate transport in a non-trivial multi-terminal geometry.
\end{abstract}


\end{frontmatter}

\section{Introduction}
\label{chap:Modeling}
For many years the workhorse of transport calculations in two-terminal ballistic systems has been the recursive Green's
function method \cite{Ferry:book1997,Drouvelis:jcp2006}. As we will show, this method is also applicable for multi-terminal systems; but at
a heavy price for the unmodified version of the method. Several schemes have been developed to make the recursive Green's
function method suitable for multi-terminal systems. These include: An optimal block-tridiagonalization scheme \cite{Wimmer:jcp2009,Duckheim:prb2010}, a scheme
utilizing the reverse Cuthill-McKee algorithm for connected graphs \cite{Mason:prb2011},
a decimation method \cite{Pastawski:arxiv2001,Gopar:prb2004}, a circular slicing scheme for a simple four-terminal cross \cite{Nikolic:sst2009,Nikolic:Chapter2010},
%
and a ``knitting'' algorithm \cite{Kazymyrenko:prb2008};
of which, the knitting algorithm has been particularly popular in the research community \cite{Akhmerov:prb2009,Chen:prb2010,Haugen:epl2011}.
Our focus, however, will be on the circular slicing scheme. This scheme is a simple modification of the original
two-terminal scheme. Also, performance wise, it is in the same caliber as other advanced methods,
such as the knitting algorithm.
In this paper we will generalize the circular slicing scheme presented in Ref.\ \cite{Nikolic:Chapter2010}.
With this generalization the circular slicing scheme is easy to use and suitable for a large variety of different
multi-terminal geometries; with minimum additional development between systems.

The paper is organized as follows. We will start in Sec.\ \ref{sec:DiscHDer} by demonstrating how to discretize a general
one particle Hamiltonian. In Sec.\ \ref{sec:RecG} we will present the original two-terminal recursive Green's function
method. How we generalize the original recursive Green's function method for multi-terminal systems via the circular slicing scheme 
is presented in Sec.\ \ref{sec:circMeth}. In Sec.\ \ref{sec:ToySys} we apply the circular slicing method to a non-trivial 
multi-terminal system with an applied magnetic field. We will discuss the performance and stability of the method in Sec.\ \ref{sec:CircPer}.
Lastly, in Sec.\ \ref{sec:CircConl} we  will present a discussion on internal degrees of freedom, mixed schemes and a comparison with the knitting
algorithm.

\section{Discretizing the system}
\label{sec:DiscHDer}
\begin{figure}[btp!]
 \centering
 \includegraphics[width=0.6\textwidth]{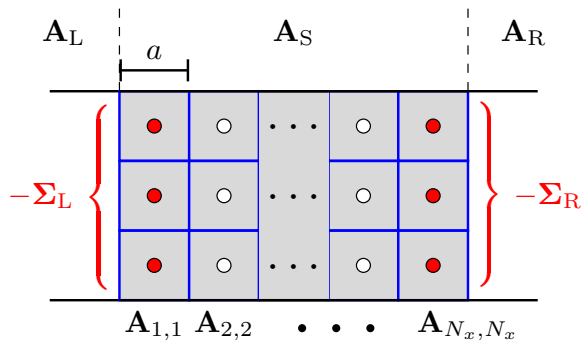}
\caption{(Color online) Schematic showing the two-terminal system discretized on a grid with mesh size $a$. The system is divided into
        three areas; a scattering region represented by the matrix $\mathbf{A}_{\mathrm{S}}$ and two semi-infinite leads represented by the
        matrices $\mathbf{A}_{\mathrm{L}}$ and $\mathbf{A}_{\mathrm{R}}$. By grouping the discretized points into vertical slices,
        $\mathbf{A}_{\mathrm{S}}$ takes the form of a block tridiagonal matrix. Slice $i$ is then represented by the block
        $\mathbf{A}_{i,i}$. The effects of the semi-infinite leads add as self-energies $\boldsymbol{\Sigma}_{\mathrm{L}}$ and
        $\boldsymbol{\Sigma}_{\mathrm{R}}$, to the points at the left and right boundaries of the scattering region.}
 \label{fig:normalSlicing}
\end{figure}

The purpose of the recursive Green's function method is to partially invert a matrix that is on a block tridiagonal form. In our case
this matrix is the Hamiltonian of the system, which we discretize using a three-point finite difference scheme \cite{Bradie:book2006}.
We start by considering a simple two-terminal system. Our focus will be on two dimensional systems; since the algorithm used here
was originally specifically designed for such systems.
Note though that the method is not limited to two dimensional systems and can readily be extended to three dimensional systems.
In Fig.\ \ref{fig:normalSlicing} we have a schematic
of the sample under study. We describe this system with the single particle Hamiltonian
\begin{equation}
H=\int\Psi^{\dagger}(\mathbf{r})H(\mathbf{r})\Psi(\mathbf{r})d\mathbf{r}
\end{equation}
where we have written it with real space field operators $\Psi(\mathbf{r})$.
We begin by discretizing the Hamiltonian on a grid with mesh size $a$. Over each mesh we assume that the field operators $\Psi(\mathbf{r})$ have
a constant value, evaluated at the center point of the mesh. This transforms the Hamiltonian to a matrix
\begin{equation}
\mathbf{H}=\sum_{i,j}\Psi^{\dagger}(i,j)H(i,j)\Psi(i,j)+O(a^2),
\end{equation}
where the $i$, and $j$ label the center points of the meshes and we have redefined the field operators as
$\Psi(\mathbf{r})\rightarrow\Psi(i,j)/a$. The goal is to find the retarded Green's function of $\mathbf{H}$.
Note that the retarded Green's function $\mathbf{G}^{\mathrm{r}}$ is just the inverse of 
\begin{eqnarray}
 \mathbf{A}=E\mathbf{I}-\mathbf{H}.
\label{eq:Amat}
\end{eqnarray}
In the discretizing procedure we use a simple three-point central 
difference scheme for gradients and nablas. For example, the $x$ part of the kinetic energy is approximated as
\begin{align}
T_x&=-\frac{\hbar^2}{2m}\int \Psi^{\dagger}(\vec{r})\left(\frac{\partial^2}{\partial x^2}\Psi(\vec{r})\right)d\vec{r}\nonumber\\
&\approx-t\sum_{i,j}\left[\Psi^{\dagger}(i,j)\Psi(i+1,j)-2\Psi^{\dagger}(i,j)\Psi(i,j)+\Psi^{\dagger}(i,j)\Psi(i-1,j)\right],
\end{align}
where $t=\hbar^2/2m^*a^2$ results from the discretization.
In all numerical calculations we scale the energy in the factor $t$, known as the tight-binding hopping parameter.
%

At this point in our discussion $\mathbf{A}$ is still an infinite matrix that describes the whole system (leads and all). We are interested in what happens in the central 
scattering region. In order to focus our attention to that region we divide the Hamiltonian for the whole system into three parts:
the left and right leads, represented by $\mathbf{A}_{\mathrm{L}}$ and $\mathbf{A}_{\mathrm{R}}$ respectively, and the scattering region, represented by $\mathbf{A}_{\mathrm{S}}$,
see Fig.\ \ref{fig:normalSlicing}. We do the same for the Green's function of whole system and write
\begin{small}
\begin{equation}
\left[\begin{array}{ccc}
                 \mathbf{A}_{\mathrm{L}}  & \mathbf{A}_{\mathrm{LS}}  & 0     \\ 
                 \mathbf{A}_{\mathrm{SL}} & \mathbf{A}_{\mathrm{S}}   & \mathbf{A}_{\mathrm{SR}}  \\
                      0                  & \mathbf{A}_{\mathrm{RS}}  & \mathbf{A}_{\mathrm{R}} \\
                \end{array}\right]
          \left[\begin{array}{lll}
                 \mathbf{G}_{\mathrm{L}}^{\mathrm{r}} & \mathbf{G}_{\mathrm{LS}}^{\mathrm{r}} & \mathbf{G}_{\mathrm{LR}}^{\mathrm{r}} \\ 
                 \mathbf{G}_{\mathrm{SL}}^{\mathrm{r}} & \mathbf{G}_{\mathrm{S}}^{\mathrm{r}} & \mathbf{G}_{\mathrm{SR}}^{\mathrm{r}} \\
                 \mathbf{G}_{\mathrm{RL}}^{\mathrm{r}} & \mathbf{G}_{\mathrm{RS}}^{\mathrm{r}} & \mathbf{G}_{\mathrm{R}}^{\mathrm{r}} \\
                \end{array}\right]
        =\left[\begin{array}{ccc}
                 \mathbf{I} & 0 & 0 \\ 
                 0 & \mathbf{I} & 0 \\
                 0 & 0 & \mathbf{I} \\
                \end{array}\right].
\label{eq:HGmatrixEq}
\end{equation}
\end{small}
\noindent The matrices $\mathbf{A}_{\mathrm{S}j}=\mathbf{A}^{\dagger}_{j\mathrm{S}}=-\mathbf{H}_{\mathrm{S}j}=t\mathbf{I}$ result from the discretization and
couple together the scattering region to the left lead ($j=$ L) and right lead ($j=$ R). Note that there is no direct coupling between the left and right leads,
i.e, $\mathbf{A}_{\mathrm{LR}}=0$ and $\mathbf{A}_{\mathrm{RL}}=0$.
Multiplying out Eq.\ (\ref{eq:HGmatrixEq}) gives nine matrix equations from which we can isolate a
\emph{finite} matrix equation for $\mathbf{G}_{\mathrm{S}}^{\mathrm{r}}$ 
\begin{equation}
\left(\mathbf{A}_{\mathrm{S}}-\sum_j \boldsymbol{\Sigma}_j\right)
     \mathbf{G}_{\mathrm{S}}^{\mathrm{r}}
 =\mathbf{I},
 \label{eq:FiniteGreenDer}
\end{equation}
whose dimension is the number of spatial points in the scattering region.
Here $\boldsymbol{\Sigma}_j=\mathbf{A}_{\mathrm{S}j}
                        \mathbf{A}_{\mathrm{j}}^{-1}
                        \mathbf{A}_{j\mathrm{S}}$
is the self-energy contribution from lead $j$. The same procedure as described above can be used for any lead connecting to the system.
Therefore, Eq.\ (\ref{eq:FiniteGreenDer}) also holds for multi-terminal systems where $j$ will run over all leads.

The leads are usually assumed to be translationally invariant.
This means that the self-energy matrix $\boldsymbol{\Sigma}_j$ can in most cases be worked out
analytically \cite{Datta:book1995} and almost always numerically via a quickly convergent iteration scheme \cite{LopezSancho:jpf1985}.
To simplify the notation we will from now on refer to $\mathbf{A}_{\mathrm{S}}$ and $\mathbf{G}_{\mathrm{S}}^{\mathrm{r}}$ as simply $\mathbf{A}$
and $\mathbf{G}^{\mathrm{r}}$; since we are only interested in the Green's function for the scattering region.
For a two-terminal system we order the matrix elements of $\mathbf{A}$ in such a way that each diagonal block $\mathbf{A}_{i,i}$ represents a vertical
slice at $i$ in our scattering region, see Fig. \ref{fig:normalSlicing}. With this ordering the self-energies from the left
and right leads add only to the first and last diagonal blocks respectively, i.e., $\mathbf{A}_{1,1}$ and  $\mathbf{A}_{N_x,N_x}$.
This is important, because the block tridiagonal form is then preserved; enabling us to use the recursive algorithm.

\section{The recursive Green's function algorithm for two-terminal systems}
\label{sec:RecG}

The recursive algorithm revolves around finding either the diagonal blocks or selected block columns of an inverted block tridiagonal matrix.
This method is sometimes called partial inversion since only part of the Greens function is calculated. For many physical quantities, only parts of the full Green's function are required to compute them,
e.g, for charge density we need the first and last columns. The recursive Green's function method therefore offers an economical approach
for the computation. 
\begin{figure}[b!]
 \centering
 \includegraphics[bb= 131 490 552 720,width=1\textwidth]{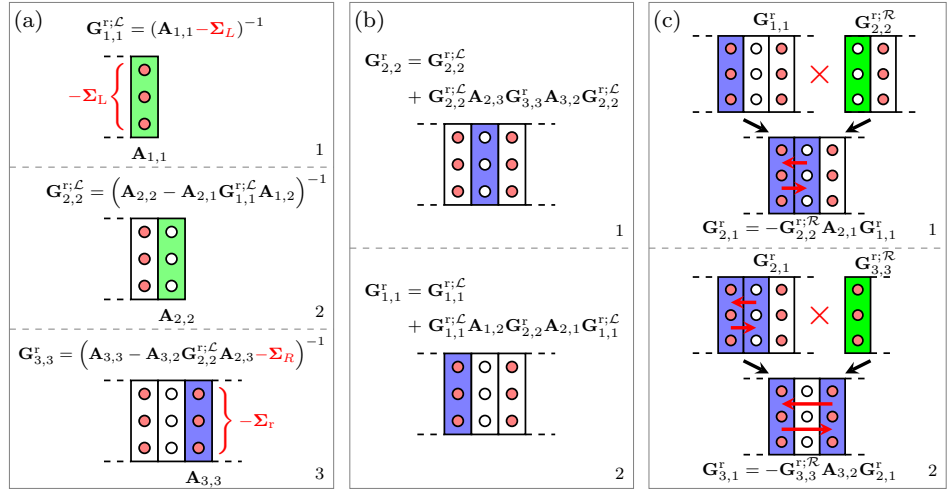}
\caption{(Color online) Schematic showing the recursive Green's function method applied to a $3\times3$ two-terminal example system. Figure (a) shows
         the forward algorithm using Eqs.\ (\ref{eq:G_Forward_A}), (\ref{eq:G_Forward_B}), and (\ref{eq:G_Forward_C}). The backward
         algorithm, using Eq.\ (\ref{eq:G_Backward}), is shown in (b). In (c) the first column of the full Green's function
         is calculated via Eq.\ (\ref{eq:G_lowerColumn}).}
 \label{fig:RecGForward}
\end{figure}
The algorithm is divided into three parts. In Fig.\ \ref{fig:RecGForward} we show schematically the workings of the algorithm
for a simple $3\times3$ example system.

(i) The first part is the so called forward algorithm and is demonstrated in Fig.\ \ref{fig:RecGForward} (a) for our example system. This
part uses the same method outlined above where we ``folded'' the effects of the
semi-infinite leads into the scattering region. We begin by dividing the system into vertical slices. The first slice, already containing the
self-energy of the left lead, is inverted and added to the next slice to the right. This procedure is continued until
we reach the last slice:
\begin{align}
 \label{eq:G_Forward_A}
 \mbox{First slice:}&\quad\mathbf{G}^{\mathrm{r;}\mathcal{L}}_{1,1}=(\mathbf{A}_{1,1}-\boldsymbol{\Sigma}_{\mathrm{L}})^{-1} \\
 \label{eq:G_Forward_B}
  \mbox{$i$-th slices:}&\quad\mathbf{G}^{\mathrm{r;}\mathcal{L}}_{i,i}=(\mathbf{A}_{i,i}-\mathbf{A}_{i,i-1}\mathbf{G}^{\mathrm{r;}\mathcal{L}}_{i-1}\mathbf{A}_{i-1,i})^{-1}, \quad i=2,\dots,N_x-1\\
 \label{eq:G_Forward_C}
 \mbox{Last slice:}&\quad\mathbf{G}^{\mathrm{r;}\mathcal{L}}_{N_x,N_x}=(\mathbf{A}_{N_x,N_x}-\mathbf{A}_{N_x,N_x-1}\mathbf{G}^{\mathrm{r;}\mathcal{L}}_{N_x-1}\mathbf{A}_{N_x-1,N_x}-\boldsymbol{\Sigma}_{\mathrm{R}})^{-1}.
\end{align}
Because the last slice contains the self-energy of the right lead, the Green's function of that slice is exact, i.e.,
$\mathbf{G}^{\mathrm{r}}_{N_x,N_x}=\mathbf{G}^{\mathrm{r;}\mathcal{L}}_{N_x,N_x}$. 
Above we applied the algorithm from the left to right and we used the notation ${\mathcal{L}}$ to mark the above Green's functions
as left-connected Green's functions. The algorithm can also be applied from the right to left; which would produce right-connected
Green's functions marked with ${\mathcal{R}}$. These right connected Green's functions will play a role in
the last part of the algorithm. 

(ii) The second part is called the backward algorithm and computes the diagonal blocks of the full Green's function. From the forward algorithm
we obtained the last block on the diagonal of the full Green's function. Using the Dyson equation we can couple it to the left connected
Green's function $\mathbf{G}^{\mathrm{r;}\mathcal{L}}_{N_x-1,N_x-1}$ of the adjacent slice on the left. This produces diagonal block number $N_x-1$
of the full Green's function. We then continue this procedure until we have calculated all the diagonal blocks:
\begin{equation}
\mathbf{G}^{\mathrm{r}}_{i,i}=\mathbf{G}^{\mathrm{r;}\mathcal{L}}_{i,i}
 + \mathbf{G}^{\mathrm{r;}\mathcal{L}}_{i,i}\mathbf{A}_{i,i+1}\mathbf{G}^{\mathrm{r}}_{i+1,i+1}\mathbf{A}_{i+1,i}\mathbf{G}^{\mathrm{r;}\mathcal{L}}_{i,i},
\quad i=1,\dots,N_x-1.
\label{eq:G_Backward}
\end{equation}
In Fig.\ \ref{fig:RecGForward} (b) we use the backward algorithm on our example system.

(iii) Lastly, the third part calculates the off-diagonal blocks. It uses the Dyson equation to couple together the exact diagonal blocks with left 
or right connected Green's functions to produce the off-diagonal blocks below or above the diagonal line.
\begin{align}
\label{eq:G_upperColumn}
\mathbf{G}^{\mathrm{r}}_{i-1,j}&=-\mathbf{G}^{\mathrm{r;}\mathcal{L}}_{i-1,i-1}\mathbf{A}_{i-1,i}\mathbf{G}^{\mathrm{r}}_{i,j},\quad 1< i\le j \le N_x \\
\label{eq:G_lowerColumn}
\mathbf{G}^{\mathrm{r}}_{i+1,j}&=-\mathbf{G}^{\mathrm{r;}\mathcal{R}}_{i+1,i+1}\mathbf{A}_{i+1,i}\mathbf{G}^{\mathrm{r}}_{i,j},\quad N_x-1\ge i\ge j \ge 1
\end{align}
In Fig.\ \ref{fig:RecGForward} (c) we apply Eq.\ (\ref{eq:G_lowerColumn}) to calculate the first column of the Green's function of our example system.

\section{Circular slicing method}
\label{sec:circMeth}
\begin{figure}[btp!]
 \centering
 \includegraphics[width=0.9\textwidth]{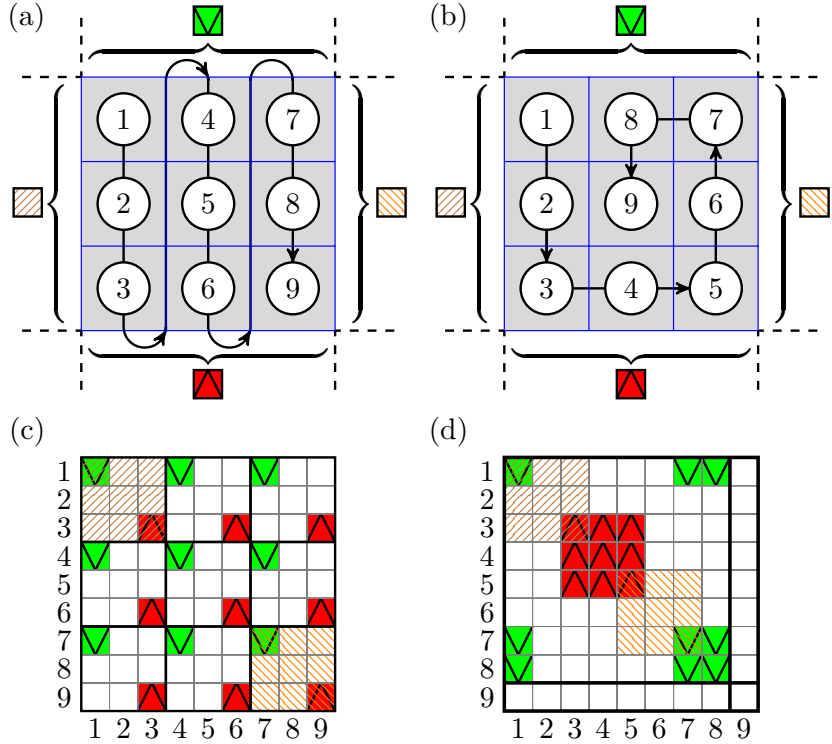}
\caption{(Color online) Schematic showing the different ordering of self-energies in a vertical slicing scheme and circular slicing for our simple $3\times3$
         example system.
         Figures (a) and (b) show the numbering of the spatial points in $\mathbf{A}$ for a vertical slicing and circular
         slicing scheme, respectively. We assume that the leads cover the whole side to which they are connected. The 
         $\boldsymbol{\Sigma}_{\mathrm{L}}$ is represented with the symbol  \protect\includegraphics[width=0.02\textwidth]{SelfLeft.eps},
         $\boldsymbol{\Sigma}_{\mathrm{R}}$ with \protect\includegraphics[width=0.02\textwidth]{Selfright.eps},
         $\boldsymbol{\Sigma}_{\mathrm{T}}$ with \protect\includegraphics[width=0.02\textwidth]{Selfup.eps},
         and $\boldsymbol{\Sigma}_{\mathrm{B}}$ with \protect\includegraphics[width=0.02\textwidth]{Selfdown.eps}.
         In (c) and (d) we see how the elements of the self-energies are 
         distributed within $\mathbf{A}$ for vertical slicing and circular sclicing, respectively.}
 \label{fig:Ordering}
\end{figure}

As noted before, we group the spatial points into vertical slices and order them along these slices. If we use this ordering within $\mathbf{A}$,
adding self-energies of the left and right leads will not destroy the block tridiagonal structure. The self-energies are contained within
in the first and last diagonal blocks. However, this is not the case if we add self-energies from leads connected at the bottom or top sides of the
scattering region. The elements from these self-energies will add to all blocks (off and on the diagonal) that are associated with
the leads connected to the bottom or top.
This is illustrated in Figs. \ref{fig:Ordering} (a) and \ref{fig:Ordering} (c) for a simple four terminal system, discretized on 
a $3\times3$ grid. Fig. \ref{fig:Ordering} (a) shows the ordering of the spatial points and where the four
self-energies, $\boldsymbol\Sigma_{\mathrm{L}}$, $\boldsymbol\Sigma_{\mathrm{R}}$, $\boldsymbol\Sigma_{\mathrm{T(op)}}$, and $\boldsymbol\Sigma_{\mathrm{B(ottom)}}$ are added.
The resulting ordering of the self-energy matrix elements into the $\mathbf{A}$ matrix is shown in Fig. \ref{fig:Ordering} (c).
Note that we use the symbols \includegraphics[width=0.02\textwidth]{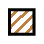}, \includegraphics[width=0.02\textwidth]{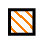},
\includegraphics[width=0.02\textwidth]{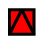}, and \includegraphics[width=0.02\textwidth]{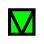} to denote
the matrix elements for $\boldsymbol{\Sigma}_{\mathrm{L}}$, $\boldsymbol{\Sigma}_{\mathrm{R}}$, $\boldsymbol{\Sigma}_{\mathrm{T}}$,
and $\boldsymbol{\Sigma}_{\mathrm{B}}$, respectively.

\begin{figure}[btp!]
 \centering
 \includegraphics[width=0.9\textwidth]{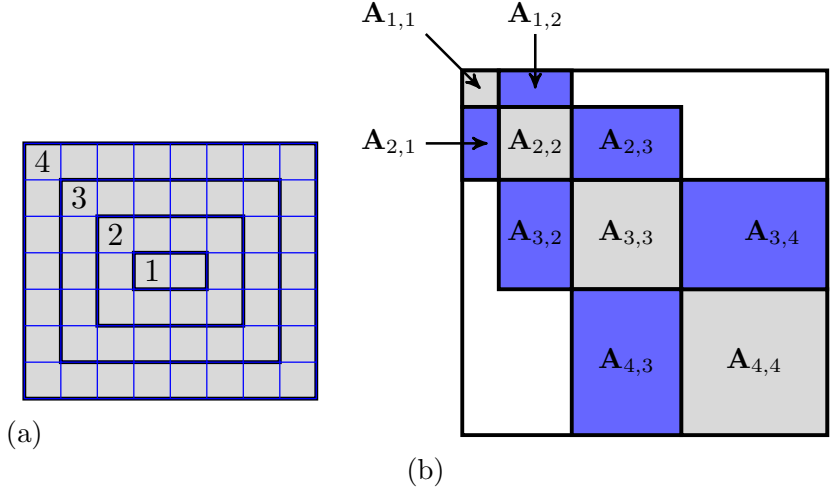}
\caption{(Color online) In (a) we show an schematic example of the rectangular grouping of spatial points. The structure of the resulting
block tri-diagonal matrix can be seen in (b).}
 \label{fig:CircMat}
\end{figure}

A quick fix would be to join all the slices connected to the bottom
and top leads into one large slice; thereby preserving the block tridiagonal structure. But this solution is computationally expensive.
A more computationally economical approach is to change the ordering of the spatial points connected to the bottom and top leads.
By grouping the points into rectangular slices instead of vertical slices the block tridiagonal structure is regained.
An example of this grouping can be seen in Fig.\ \ref{fig:CircMat} (a).
For demonstration, we look again at the simple four terminal system discretized on a $3\times3$ grid. The left and right leads
are of width $N_y$ and the top and bottom transverse leads are of width $N_{\mathrm{T(ransverse)}}$. 
Fig.\ \ref{fig:Ordering} (b) shows the circular ordering of the spatial points. Now, instead of ordering in vertical slices
from top to bottom we order the points in a rectangle shaped spiral. The result is that all the matrix elements of the self-energies
end up in one block on the diagonal of $\mathbf{A}$. This block represents the outermost rectangle, see Fig. \ref{fig:Ordering} (d).
The outermost rectangle is the largest, with $2N_y+2N_{\mathrm{T}}-4$ points. The next rectangle slice contains 8 fewer points and so on; until
the innermost slice is reached.
Interestingly, the block matrix that contains the self-energies now grows as the surface of the system instead of the total area of the system,
i.e., as $\sim N_y$ but not as $\sim N_y^2$ (if we assume $N_y=N_{\mathrm{T}}$).
Note that in the circular slicing scheme the nearest neighbor coupling between points is still the same as for the vertical slicing scheme.
All that we are doing is shuffling rows and columns in the original $\mathbf{A}$ matrix. As an example, let's consider the system presented in
Fig.\ \ref{fig:CircMat} (a). For that system the resulting block tridiagonal structure of $\mathbf{A}$, produced by the circular slicing scheme,
would be similar to what we see in Fig.\ \ref{fig:CircMat} (b).

In the circular slicing scheme each of the blocks on the diagonal $\mathbf{A}_{i,i}$ is almost tridiagonal.
The only departure from the tridiagonal form are two matrix elements on upper right
and lower left corners of the blocks. 
These outlying matrix elements are due to the circular form of the slices; reminiscent of periodic boundary conditions.
Note that  if $N_y\ne N_{\mathrm{T}}$ the block that corresponds to the innermost slice is an exception. This is due to the possibility of
more connections between internal points, which in turn results in a more significant departure from the tridiagonal form.

The off-diagonal blocks $\mathbf{A}_{i,i\pm 1}$, that couple together the rectangular slices, have highly symmetrical structure.
They only deviate slightly from the diagonal form of the coupling matrices that result from the standard vertical slicing scheme.
\section{Multi-terminal system}
\label{sec:ToySys}
\begin{figure}[tbp!]
 \centering
 \includegraphics[width=0.8\textwidth]{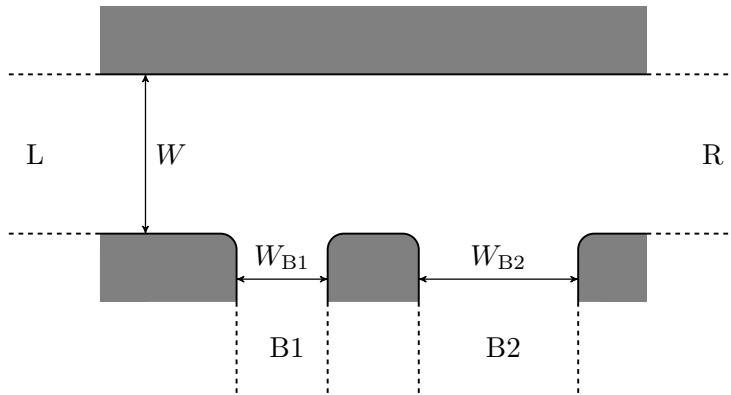}
\caption{Schematic of the example system. It is a $L_{\mathrm{S}}=686$ nm long
and $W_{\mathrm{S}}=286$ nm wide four terminal system. At the left and right side it is connected to $W=200$ nm
wide leads, labeled L and R respectively. At the bottom it is also connected to two leads labeled B1 and B2. 
Lead B1 is $W_{\mathrm{B1}}=114$ nm wide and lead B2 is $W_{\mathrm{B2}}=200$ nm wide. All corners are rounded and
the whole system, leads and scattering region, is subjected to the same magnetic field $B$ oriented perpendicular to the 2DEG. }
 \label{fig:ExSysScehma}
\end{figure}

We will now outline how the circular slicing method can be applied to a non-trivial multi-terminal setup.
As an example system let's consider the setup shown and described in Fig.\ \ref{fig:ExSysScehma}. 
We are interested in the conductance between
different hard-walled leads and the charge density of the system. To highlight the multi-terminal character we will consider
injection of electrons from left lead L into the sample and examine the behavior of the system as a function of the magnetic field $B$; which
is perpendicular to the 2DEG. 
Because the leads contain a magnetic field $B$, we calculate their self-energy contributions via a highly convergent iteration method
\cite{LopezSancho:jpf1985}.
The magnetic field is incorporated into the discretized Hamiltonians via a Peierls substitution.
 
The electrons are injected with energy $\mu_{\mathrm{F}}=8 E_0$, where
\begin{equation}
E_0=\pi^2\hbar^2/2m^*W^2=5.76\times10^{-2}t
\end{equation}
is the lowest
transverse energy of lead L, R and B2. Therefore, the injected electrons can occupy the lowest two energy bands of lead L, R and B2.
Lead B1, however, is narrower than the other leads. This means that its energy bands are at higher energy.
The injected electrons can therefore only enter the lowest mode of lead B1.
 We discretize the system on a $120\times50$-point grid.
When calculating the necessary Green's function elements we will group the spatial points into rectangular slices, similar to those shown
in Fig.\ \ref{fig:CircMat} (a). 

\subsection{Conductance}
\label{sec:Cond}
To calculate the conductance we will use the Landauer-B\"{u}ttiker formalism and assume that we are in the linear response regime 
\cite{Datta:book1995,Blanter:PhysRep2000,Baranger:prb1989}.
The conductance from the left lead L to a lead $j$ ($j=$R, B1, B2) for energy $\mu_{\mathrm{F}}$ is then written as 
\begin{equation}
G_{j,\mathrm{L}}(E_{\mathrm{F}})=\frac{2e^2}{h}
\mbox{Tr}\left[\boldsymbol{\Gamma}_j \mathbf{G}^{\mathrm{r}}\boldsymbol{\Gamma}_{\mathrm{L}}\mathbf{G}^{\mathrm{a}}\right].
\label{eq:Conductance}
\end{equation}
Here $\mathbf{G}^{\mathrm{a}}=\left(\mathbf{G}^{\mathrm{r}}\right)^{\dagger}$ and
$\boldsymbol{\Gamma}_j=i\left(\boldsymbol{\Sigma}_j^{\mathrm{R}}-\boldsymbol{\Sigma}_j^{\mathrm{A}}\right)$
where $\boldsymbol{\Sigma}_j^{\mathrm{a}}=\left(\boldsymbol{\Sigma}_j^{\mathrm{r}}\right)^{\dagger}$. 

From Eq.\ (\ref{eq:Conductance}) we see why the recursive Green's function method is so convenient.
Let's assume that lead $j$ is connected to some group of spatial points A and lead L to some group of spatial points B. Note that these points are on the boundary of
the scattering region. The $\boldsymbol{\Gamma}_j$ and the $\boldsymbol{\Gamma}_{\mathrm{L}}$ matrices will then be zero for all points other than
those in A or B, respectively. This allows us to write the trace of Eq.\ \ref{eq:Conductance} as
\begin{equation}
\sum_{a,a'\in A}\sum_{b,b'\in B}
\left(\boldsymbol{\Gamma}_j\right)_{a,a'} \left(\mathbf{G}^{\mathrm{r}}\right)_{a',b}\left(\boldsymbol{\Gamma}_{\mathrm{L}}\right)_{b,b'}\left(\mathbf{G}^{\mathrm{a}}\right)_{b',a},
\label{eq:MatTrace}
\end{equation}
i.e., we only require the matrix elements of $\mathbf{G}^{\mathrm{r}}$ and $\mathbf{G}^{\mathrm{a}}$ that correspond to points in A or B. As noted before,
the points in A and B are on the boundary of the scattering region, i.e., the outermost slice.
Therefore, we only need the the diagonal block of the full Green's function that represents the outermost slice, corresponding to $A_{4,4}$ in Fig.\ \ref{fig:CircMat}, to calculate the conductance.
This is readily done with the circular slicing version of the recursive Green's function method. Furthermore,
we only need to perform the forward algorithm if we begin the algorithm on the innermost slice; because then we end up with the full Green's function
of the outermost slice.

\begin{figure}[htp!]
 \centering
 \includegraphics[width=0.8\textwidth]{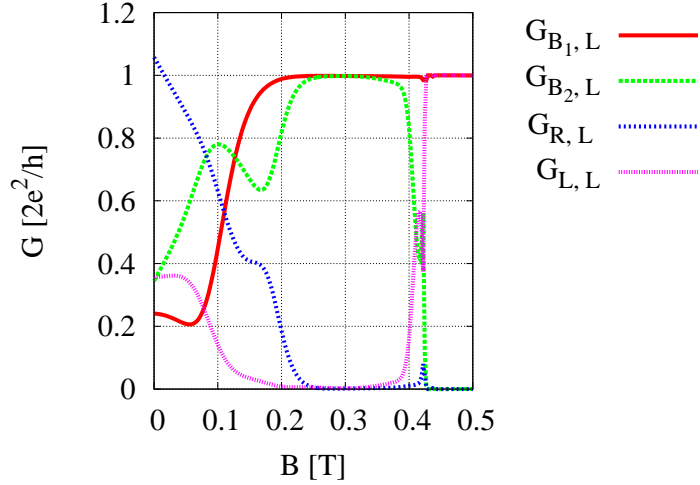}
\caption{(Color online) Conductance from the left lead L to other leads, B1, B2 and R, and backscattering to itself as a function of magnetic field.}
 \label{fig:ContEx}
\end{figure}

Below, in Fig.\ \ref{fig:ContEx}, we present the conductance from lead L to B1,  L to B2,
and L to R, i.e. $G_{\mathrm{B1},\mathrm{L}}$, $G_{\mathrm{B2},\mathrm{L}}$, and $G_{\mathrm{R},\mathrm{L}}$, respectively.
Also in Fig.\ \ref{fig:ContEx}, we plot the backscattering into lead L defined as
\begin{equation}
G_{\mathrm{L},\mathrm{L}}=N-G_{\mathrm{B1},\mathrm{L}}-G_{\mathrm{B2},\mathrm{L}}-G_{\mathrm{R},\mathrm{L}},
\end{equation}
for electrons injected into the system in the lowest $N$ modes of lead L. In our case $N=2$. 

In Fig.\ \ref{fig:ContEx} we see that at zero magnetic fields the biggest conductance is from the left to right lead $G_{\mathrm{R},\mathrm{L}}$.
Small conductance from the left lead to the two bottom leads, $G_{\mathrm{B1},\mathrm{L}}$ and $G_{\mathrm{B2},\mathrm{L}}$, results from the scattering
with the corners. As expected, when the magnetic field increases, $G_{\mathrm{B1},\mathrm{L}}$ and $G_{\mathrm{B2},\mathrm{L}}$ do as well while
$G_{\mathrm{R},\mathrm{L}}$ decreases. This is due to the Lorentz force which pulls the electrons toward the lower end. But for low magnetic fields $B<0.1$,
$G_{\mathrm{B2},\mathrm{L}}$ dominates over $G_{\mathrm{B1},\mathrm{L}}$. This is because for low magnetic fields the electrons have a
large cyclotron radius. For $B>0.1$ the cyclotron radius is small enough to steer
the electrons increasingly toward the B1 lead. Now, because the B1 lead is narrower than the other leads, the electrons can only enter its first transverse
mode. 
The conductance $G_{\mathrm{B1},\mathrm{L}}$ therefore saturates at $2e^2/h$ around $B=0.2$ T. At this high magnetic fields the electron occupy edge
states, i.e., follow the wire walls. This means that all the electrons which do not enter lead B1, i.e. scatter of lead B1, creep into lead B2.
When the strength of the magnetic field reaches $B=0.4$ T the lowest Landau level has become so costly in energy, that the electrons which had only enough energy to enter B2 can
not propagate at all through the system. Therefore, the sum of the conductance to all the leads plummets down to $2e^2/h$.

\subsection{Density}
\label{sec:Den}

\begin{figure}[btp!]
 \centering
 \includegraphics[width=0.9\textwidth]{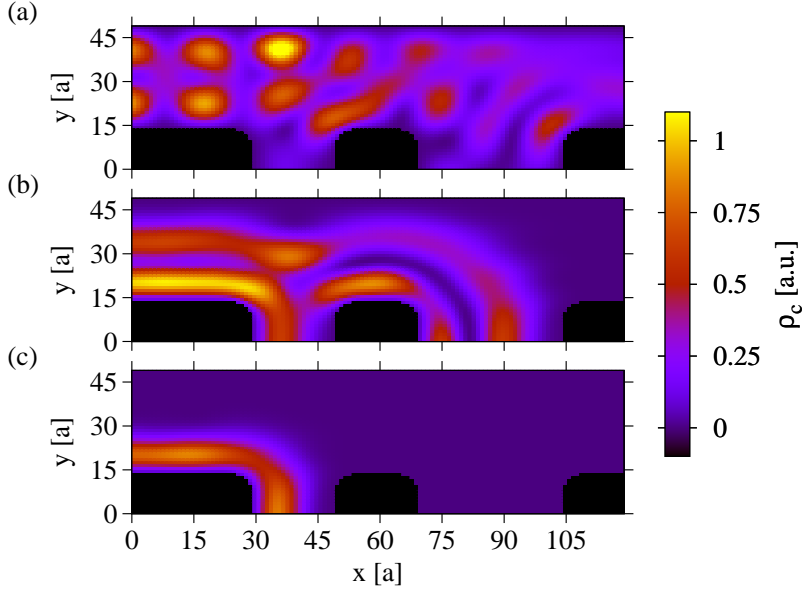}
\caption{(Color online) Charge density in the system with electrons injected from the left lead with energy $\mu_{\mathrm{F}}$. In (a) there 
         is no magnetic field while in (b) and (c) the magnetic field is set at $B=0.3$ T and $B=0.5$ T, respectively.
         Note that the charge density is presented in arbitrary units.}
 \label{fig:DensEx}
\end{figure}

Via the Keldysh formalism \cite{Keldysh:JETP1965,Haug:book2008} we can calculate the charge density of the system as
\begin{equation}
\rho_c(i,j)=
e\int_0^{E_F}\frac{dE}{2\pi}\sum_n\mathbf{G}^{\mathrm{r}}\left[\boldsymbol{\Gamma}_n f_n(E)\right]\mathbf{G}^{\mathrm{a}}\Bigg|_{(i,i),(j,j)}
\label{eq:TotDens}
\end{equation}
%
%
where $f_n(E)=1/[\exp(\beta(E-\mu_j))+1]$ is the Fermi function in lead $n$ which has the chemical potential
$\mu_n=E_{\mathrm{F}}+eV_j$. Considering the matrix multiplication of $\boldsymbol{\Gamma}_n$ with $\mathbf{G}^{\mathrm{r}}$ and $\mathbf{G}^{\mathrm{a}}$
in Eq.\ \ref{eq:TotDens} we note that only the last column of the full Green's function is needed. Looking at Fig.\ \ref{fig:CircMat} (b) 
we see that the last column is readily calculated by following all three steps of the recursive Green's function algorithm described in Sec.\ \ref{sec:RecG}.
We would like to corroborate the conductance results in Sec.\ \ref{sec:Cond} with the corresponding charge density calculation. Using Eq.\ (\ref{eq:TotDens}) is not the best
way to do this because Eq.\ (\ref{eq:TotDens}) gives us the density information for the whole energy range. We would rather want to get information 
about the density contribution from an small energy interval around $\mu_{\mathrm{F}}$. Therefore, we assume linear response and replace the Fermi energy $f_n(E)$ with
a delta function
$\delta(E-\mu_{\mathrm{F}})$, i.e.,
\begin{equation}
\rho_c(i,j)\big|_{E=\mu_{\mathrm{F}}}=
\frac{e}{2\pi}\sum_n\mathbf{G}^{\mathrm{r}}\boldsymbol{\Gamma}_n\mathbf{G}^{\mathrm{a}}\Bigg|_{(i,i),(j,j)}.
\label{eq:TotDens_muF}
\end{equation}
In Fig.\ \ref{fig:DensEx} we calculate via Eq.\ (\ref{eq:TotDens_muF}) the density resulting from an injection of electrons with energy $\mu_{\mathrm{F}}$.
We consider three cases of magnetic field strengths: $B=0$ T, $B=0.3$ T, and $B=0.5$ T. These cases are portrayed in Fig.\ \ref{fig:DensEx} (a), (b) and (c),
respectively.

These cases corroborate well our interpretation of the conductance result. In Fig.\ \ref{fig:DensEx} (a) we see a large influence of the corners
that scatter the electrons mostly into the left and right leads. For $B=0.3$ T, shown in Fig.\ \ref{fig:DensEx} (b), we see the case where
$G_{\mathrm{B1},\mathrm{L}}$ is saturated. As seen in the conductance results the electrons that scatter off lead B1 creep along the walls into lead B2.
Lastly, for the case where $B=0.5$ T shown in Fig.\ \ref{fig:DensEx} (c), only the lowest Landau level is accessible and all the electrons end up 
in the B1 lead.

\section{Performance of the circular slicing scheme}
\label{sec:CircPer}
\subsection{Operations count}
When calculating the conductance in the circular slicing scheme, only the forward algorithm is required. This is because the forward algorithm returns the last
diagonal block of the full Green's function; which is required to calculate the conductance, see Sec.\ \ref{sec:Cond}. With this
in mind, let's compare the operation count of the circular slicing scheme to the standard vertical slicing scheme.
For the standard two-terminal system of length $N_{\mathrm{T}}$ and width $N_y$ the operation count for the forward algorithm
is $\#op_{\mathrm{2T(erm)}}\sim N_{\mathrm{T}}N_y^3$. The reason for this is simple, we are inverting $N_{\mathrm{T}}$ matrices of size $N_y\times N_y$.
Let's assume that we add a top and/or bottom lead of width $N_{\mathrm{T}}$ to the system and stick with the vertical slicing scheme.
To preserve the block tridiagonal form of $\mathbf{A}$ we will have to join the $N_{\mathrm{T}}$ slices connected to the transverse leads.
This means, again if we stick to the vertical slicing scheme, that we have to perform a direct diagonalization (brute force inversion) of a
$N_{\mathrm{T}}N_y\times N_{\mathrm{T}}N_y$ matrix. The operation count of the direct diagonalization is 
$\#op_{\mathrm{M(ulti)T(erm)}}\sim (N_{\mathrm{T}}N_y)^3=N_{\mathrm{T}}^2(\#op_{\mathrm{2T}})$. Therefore, the ratio of required operations
of the multi-terminal and two-terminal systems, grows quadratically with the width of the transverse leads.

Let's now examine the situation if we recover the block tridiagonal structure by converting to the circular slicing scheme.
If we assume $N_{\mathrm{T}}=N_y$ the number of required operations now drops down to
$\#op_{\mathrm{MTC(ircular)}}\sim 8N_{\mathrm{T}}N_y^3=8 (\#op_{\mathrm{2T}}$), i.e.,
to the same order of magnitude as the normal vertical slicing scheme for the two-terminal system.
This operation count is found by summing the number of operations required to invert the increasingly larger circular slices; from 
the innermost $4\times4$ point slice to the outermost $(4N_y-4)\times(4N_y-4)$ slice (assuming $N_{\mathrm{T}}=N_y$).
If $N_{\mathrm{T}}< N_y$ the number of operations
diminishes even further.

\begin{figure}[tbp!]
 \centering
 \includegraphics[width=0.8\textwidth]{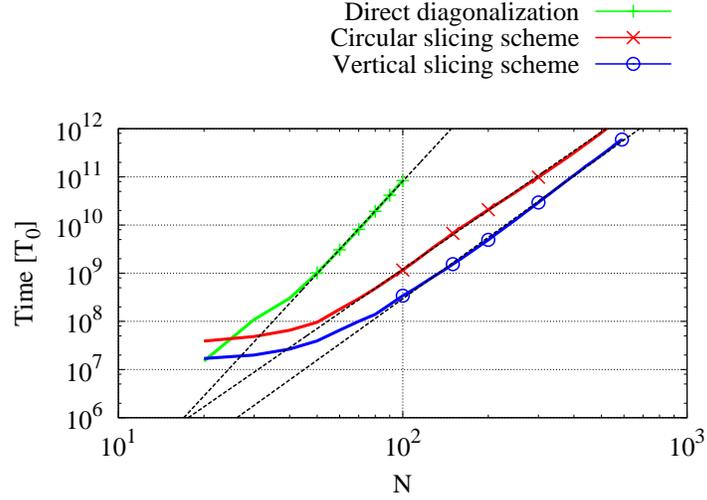}
\caption{(Color online) Comparison of the time taken to calculate one conductance point by:
         the circular slicing scheme, vertical slicing scheme, and by direct diagonalization.
         Note that the system is box shaped, i.e., $N_x=N_y=N$ and that vertical slicing scheme
         was applied to a two-terminal system while the circular slicing scheme and the direct diagonalization 
         was applied to a multi-terminal system.
         The dashed lines are fits to the points marked on the curves using Eq.\ \ref{eq:TimePow}.
         The resulting fitted parameters are shown in Table \ref{tab:Fitting}. }
 \label{fig:Timing}
\end{figure}

In Fig.\ \ref{fig:Timing} we present the time required to calculate the conductance through a two-terminal box shaped system ($N_x=N_y=N$) via three
methods: the vertical slicing scheme, the circular slicing scheme, and by direct diagonalization.
Note that for the circular slicing scheme and the direct diagonalization the system is connected to vertical leads, i.e., $\Sigma_{\mathrm{T}}\neq0$
and $\Sigma_{\mathrm{B}}\neq0$, while for the vertical slicing scheme we only include left and right leads,
i.e., $\Sigma_{\mathrm{T}}=0$ and $\Sigma_{\mathrm{B}}=0$. This is because the vertical slicing scheme can only handle two-terminal systems, while
circular slicing scheme and direct diagonalization can handle both two-terminal and multi-terminal systems.
We pick a few points on each curve and make a fit to the equation
\begin{equation}
T=T_0N^a.
\label{eq:TimePow}
\end{equation}
The fitted parameters for each scenario are shown in Table \ref{tab:Fitting}, where we scale 
the time in the $T_0$ parameter for the forward algorithm of the vertical slicing scheme. By examining the values of the fitted parameters we see that
they correspond well to the operation counts discussed above, i.e, the difference between the time required by the circular slicing scheme and the vertical
slicing scheme is a factor 8.
\begin{table}[tbp!]
\centering
\begin{tabular}{l|cc}
                        & $T_0$ & $a$\\
\hline
Vertical slicing scheme  & 1 & 4.2\\
(only forward algorithm) &  & \\
\hline
Circular slicing scheme & 8.2 & 4.1\\
\hline
Direct diagonalization & 0.014 & 6.4\\
\end{tabular}
\caption{The fitted parameters for the marked points in Fig.\ \ref{fig:Timing} using Eq.\ \ref{eq:TimePow}.}
\label{tab:Fitting}
\end{table}

\subsection{Memory requirements}
To complete the forward algorithm, the lowest amount of memory required scales as $\sim (2N_{\mathrm{T}}+2N_y-4)^2$ or $\sim 16 N_y^2$ if we assume $N_{\mathrm{T}}=N_y$
, i.e., the size of the diagonal block matrix for the largest slice. This minimum amount of required memory is achievable if we can, after each iteration, freely discard old left connected Green's functions.
This is the case if we are calculating the conductance through the whole system; in which case we only need the last diagonal block of the Green's function.
The last diagonal block represents the outermost slice and contains the self-energy contribution from all the leads 
But for cases where we have to use the backward algorithm, e.g., if we want calculate the local density, we need to store all of the left connected 
Green's functions. This amounts to $N_{\mathrm{T}}/2$ left connected Green's functions (if $N_{\mathrm{T}}$ is a even number). Therefore,
by summing the left connected Green's functions the required memory for the
backward algorithm scales as $\sim 2/3 N_y^3$, if we again assume that $N_{\mathrm{T}}=N_y$.


\subsection{Numerical stability}
On the subject of stability, a word of caution is in order. The circular slicing scheme fails if $\det({\mathbf{A}_{i,i}})=0$.
This happens if we begin the forward algorithm with the innermost slice and set the energy $E$ to be an eigenenergy of
the Hamiltonian $\mathbf{\mathbf{H}}_{i,i}$ of slice $i$.
However, the eigenenergies in question are larger or equal to the tight-binding hopping parameter $t$. Thus by ensuring that $E\ll t$, which is also the requirement for a good numerical accuracy,
the algorithm works as expected. This corresponds to the situation when $L\gg a$ where $L$ is the length of the system.

The vertical slicing scheme is not vulnerable to this kind of instability. The reason is that we begin by adding a complex self-energy matrix to the first slice.
Therefore, all eigenenergies of the Hamiltonians involved are complex and the situation $\det({\mathbf{A}_{i,i}})=0$ never occurs because the energy $E$ is real.
Interestingly, this reasoning also applies for the circular slicing scheme if we begin the forward algorithm with the outermost slice;
which unlike the innermost slice contains complex self-energy contributions.
This, however, will give the \emph{first} diagonal block of the Green's function, i.e., the block that represents the innermost slice,
see Fig.\ \ref{fig:CircMat}. For the conductance calculations we need the \emph{last} diagonal block of the Green's function, i.e.,
the block that represents the outermost slice. Therefore, to avoid using the backward algorithm and save computation time, we choose 
to start the forward algorithm with the innermost slice.

\section{Discussion}
\label{sec:CircConl}
\subsection{Including internal degrees of freedom}
As we have presented the recursive Green's function method above, we have simplified our approach to spatial points without any internal degrees
of freedom such as spin. Including such internal degrees of freedom is easily done. In our discussion above, each of the matrix elements in
$\mathbf{A}$ and $\mathbf{G}^{\mathrm{r}}$ corresponded to two spatial points, $(i,j)$ and $(i',j')$. To include the desired internal
degree of freedom,
one simply replaces each matrix element with a submatrix that describes the internal degree of freedom.
The circular slicing scheme is still perfectly applicable despite this replacement.
One only has to remember that with the circular slicing scheme we are only shuffling the labels of spatial points. We therefore keep the submatrices
intact while we shuffle them around with corresponding spatial points.

\subsection{Mixed scheme of circular and vertical slicing}
\begin{figure}[btp!]
 \centering
 \includegraphics[width=0.8\textwidth]{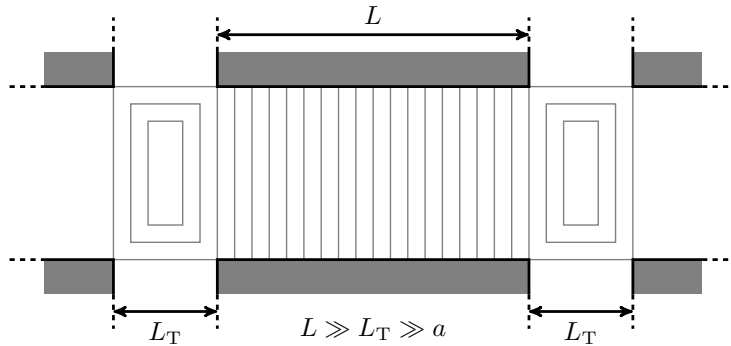}
\caption{An example of a multi-terminal system where a mixed scheme of circular and vertical slices would be appropriate. Here $a$ is the 
mesh size of the discretized grid.}
 \label{fig:MultiMulti}
\end{figure}

For systems where the transverse leads are far apart, the optimal performance of the circular slicing scheme is achieved if we use it alongside the
original vertical slicing scheme. This is done by using the circular slicing scheme only on the slices that connect to transverse
leads; using the original two-terminal scheme on the rest. This is easy to implement because both schemes use essentially the same
algorithm. We only have to shuffle the lines and columns of the matrix $\mathbf{A}$ that correspond to the transverse leads. An example
of an system where this would be a suitable approach is seen in Fig.\ \ref{fig:MultiMulti}

\subsection{Comparison with the knitting algorithm}
The circular slicing scheme is not the only way to generalize the recursive Green's function method for complex multi-terminal systems. One such generalization is
the knitting algorithm \cite{Kazymyrenko:prb2008}. The knitting algorithm takes the grouping of the grid points to the lowest extreme and adds one point at a time
in the forward and backward
algorithm; similar to knitting a sweater as the name suggests. In regard to required number of operations and memory, the method is \emph{on par} with the circular slicing
scheme. In the forward algorithm of the knitting algorithm, required operations and memory scale as $\sim N_{\mathrm{T}}N_y^2$  and $\sim N_y^2$, respectively. For the backward
algorithm of the knitting scheme, the required memory scales as $\sim N_{\mathrm{T}}N_y^2$. If we compare this to numbers in Sec.\ \ref{sec:CircPer},
we see that the performance of the two algorithms scales equally. The main
difference lies in how much the knitting algorithm and the circular slicing scheme deviate from the original two-terminal algorithm.
In the knitting algorithm, we need to replace the original equations of the forward and backward algorithms with more complicated ones. 
For the circular slicing scheme, we can use the same set of equations as in the original algorithm. 
The only difference is that we reorder lines and columns to restore the block tridiagonal structure of the matrix $\mathbf{A}$.
  
\section*{Acknowledgement}
We gratefully acknowledge helpful discussions with Sven Th.\ Sigurdsson.
This work was supported by the Icelandic Science and Technology Research
Program for Postgenomic Biomedicine, Nanoscience and Nanotechnology, the Icelandic Research Fund, and the Research
Fund of the University of Iceland.








\end{document}